\documentstyle[prl,epsfig,aps,floats]{revtex}

\newcommand\la{\langle}
\newcommand\ra{\rangle}
\newcommand\scro{{\cal O}}
\newcommand{\beq}{\begin{equation}}
\newcommand{\eeq}{\end{equation}}
\newcommand{\be}{\begin{equation}}
\newcommand{\ee}{\end{equation}}
\newcommand{\beqn}{\begin{eqnarray}}
\newcommand{\eeqn}{\end{eqnarray}} 
\def\reff#1{(\ref{#1})}
\begin{document}

\twocolumn[\hsize\textwidth\columnwidth\hsize\csname@twocolumnfalse%
\endcsname

\draft
\title{Universal Finite Size Scaling Functions in the $3D$ Ising Spin Glass}
\author{Matteo Palassini}  
\address{Department of Physics, University of California, Santa Cruz, California 95064\\
Scuola Normale Superiore and INFM, 56100 Pisa, Italy}
\author{Sergio Caracciolo}
\address{Scuola Normale Superiore and INFN, 56100 Pisa, Italy}
\date{Submitted February 10, 1999}
\maketitle

\begin{abstract}

We study the three--dimensional Edwards--Anderson model with binary interactions by 
Monte Carlo simulations. Direct evidence of finite--size scaling is
provided, and the universal finite--size scaling functions are 
determined. Monte Carlo data are extrapolated to infinite 
volume with an iterative procedure up to correlation lengths 
$\xi \approx 140$. The infinite volume data are consistent with a 
conventional power law singularity at finite temperature $T_c$. Taking into account 
corrections to scaling, we find $T_c = 1.156 \pm 0.015, \nu = 1.8 \pm 0.2$ 
and $\eta = -0.26 \pm 0.04$.
The data are also consistent with an exponential singularity at finite $T_c$, but
not with an exponential singularity at zero temperature.
\end{abstract}

\pacs{PACS numbers: 75.10.Nr, 64.60.Fr, 75.40.Mg, 75.50.Lk}
\vskip 0.1 truein
]

The critical properties of the Ising spin glass in three dimensions
are still not very well understood. Numerical simulations have led to some
progress \cite{BY,review}, but have been hampered by technical difficulties.
Large--scale Monte Carlo (MC) simulations at 
correlation length $\xi \approx 10$ lattice units 
\cite{ogielski-morgenstern,ogielski,on3D} are 
consistent with both a continuous
phase transition with power--law divergence of $\xi$ at  finite
temperature $T = T_c$, and an exponential divergence at 
$T=0$, which is expected at the lower critical dimension. 
High--statistics MC simulations of smaller systems 
\cite{KY,maparu,janke} give a certain evidence of a $T_c\neq 0$ 
transition with an ordered spin glass phase below $T_c$, but 
cannot exclude neither an exponential divergence at $T=0$, 
nor a line of critical points at $T \le T_c \neq 0$ \cite{KY,janke,iniguez}, 
as in the Kosterlitz--Thouless theory of the $2D$ XY model. Understanding 
whether an ordered spin glass phase exists in three dimensions
is clearly an issue of major interest.

In this work, we study the $3D$ Ising spin glass
with an approach, based on finite--size scaling 
(FSS) and MC si\-mulations in the paramagnetic phase, introduced in 
Ref.~\cite{fss_greedy} (see Ref.~\cite{Luscher_91} for similar methods)
and so far applied to non--disordered systems.
Let us summarize our main results.
(i) We provide a direct test of the FSS hypothesis, independent
of the nature of the divergence in the infinite system. 
In particular we determine, for the first time to our knowledge, 
the {\em universal} FSS functions. (ii) We demonstrate the effectiveness of
an iterative procedure to extrapolate the MC data to 
infinite volume, that allows us to reach $\xi \approx 140$. 
(iii) Exploiting the higher range of $\xi$, we show that an exponential 
divergence at $T=0$ is excluded, but we still cannot decide between 
a power--law divergence at $T_c \neq 0$ and a line of critical points terminating at 
$T_c\neq 0$. (iv) Under the hypothesis of power--law divergence, we 
show that corrections to scaling are important and we 
estimate $T_c$ and the critical exponents.

\paragraph*{Model and FSS method ---} 
We consider the $3D$ Edwards--Anderson model, whose Hamiltonian is
\begin{equation}
{\cal H}=- \sum_{\la xy \ra}\sigma_x J_{xy}\sigma_y  \label{model}
\end{equation}
where $\sigma_x$ are Ising spins on a simple cubic lattice of linear size $L$ with 
periodic boundaries, and $J_{xy}$ are independent random interactions
taking the values $\pm 1$ with probabi\-lity ${1\over 2}$.
The sum runs over pairs of nearest neighbor sites.

Let $\xi(T,L)$ be a suitably defined finite--volume correlation length,
and let $\scro(T,L)$ be any singular observable, such as $\xi(T,L)$ itself
or the spin--glass susceptibility (see below). Then FSS theory
\cite{Barber_FSS_review}
predicts that
\be
   {\scro(T,L) \over \scro(T,\infty)}   \;=\;
   f_{\scro} \Bigl( \xi(T,\infty)/L \Bigr) \, 
   \;,
 \label{eq1}
\ee
where $f_{\scro}$ is a universal function and corrections
to FSS are neglected. From Eq.~\reff{eq1} one obtains the relation
\beq
   {\scro(T,2L) \over \scro(T,L)}   \;=\;
   F_{\scro} \Bigl( \xi(T,L)/L \, \Bigr) \, 
   \;,
 \label{eq2}
\eeq
where $F_\scro$ is another universal function and only finite--volume observables 
are involved.
Our approach works as follows (see Ref.~\cite{fss_greedy} for
details).
We make MC runs at numerous pairs $(T,L)$, $(T,2L)$ and we
plot $\scro(T,2L) / \scro(T,L)$ versus $\xi(T,L)/L$.
If all these points fall with good accuracy on a single curve 
--- thus verifying the Ansatz (\ref{eq2}) ---
we choose a smooth fitting function $F_{\scro}$.
Then, using the functions $F_\xi$ and $F_\scro$,
we extrapolate the pair $(\xi,\scro)$ iteratively from
$L \to 2L \to 2^2 L \to \ldots \to \infty$.

\paragraph*{Computational details ---}
We simulate the model in Eq.\reff{model} 
with the heath--bath algorithm. We measure $q_x = \sigma_x \tau_x$  
and $q = L^{-3}\sum_x q_x$ from two 
independent replicas $(\sigma,\tau)$  with the same ${J_{xy}}$. 
We choose as a definition of $\xi(T,L)$ the {\em second-moment} correlation length 
\beq
  \xi(T,L)   =    
{\left[ \, S(0)/S(p) \,-\, 1 \, \right]^{1/2} \over  2 \sin(|p|/2)}
 \label{corr_len_2mom}
\eeq
where $S(k)$ is the Fourier transform 
\beq
S(k) = \sum_r e^{ik\cdot r} \la q_x q_{x+r} \ra \;,
\eeq
(arguments $T,L$ are omitted) and $p = (0,0,2\pi/L)$ 
is the smallest non--zero wave vector
\cite{notadef}. The spin--glass susceptibility is 
$\chi_{SG}(T,L) \equiv L^3 \la q^2 \ra = S(0)$.
The symbol $\la \cdot \ra$ represents a double average
over thermal noise and ${J_{xy}}$, which is estimated from
$N_s$ samples with different ${J_{xy}}$.

The runs are done on a Cray T3E parallel computer with a fast code that 
exploits the parallelism of spin glass simulations. 
The binary variables $\sigma_x$ and $J_{xy}$ at corresponding sites of 
64 samples (each represented by a single bit) are stored in a 
64-bit integer variable, and 64 $\sigma_x$'s are updated simultaneously with 
only 31 logical instructions and one random number \cite{multispin}.
Average speed on a single processor (PE) 
is  $4.5\times 10^7$ spin updates per second (DEC Alpha EV5, 600 MHz).
The PEs are arranged in a virtual parallelepiped 
along whose axis we can distribute
independent groups of 64 samples, 
 different ``slices'' of a large lattice,
and different temperatures. We typically used 32 to 128 PEs.
Equilibration of the runs is verified with the criterion introduced
in Ref.~\cite{oldBY}. The sizes simulated range 
from $L=4$ to $L=48$, from which we form 104 pairs $(T,L)$, $(T,2L)$. 
In Table I some parameters of the simulations are given.
The equivalent of about 2 years of computer time on a single PE was employed.

\begin{table}[h]
\caption{Maximum number of samples $N_s$, minimum temperature $T_{m}$ and
Monte Carlo sweeps (MCS) performed at $T_m$, as a function of the size $L$. }
\begin{center}
\begin{tabular}{lrrrrrrrrr}
$L$   & 4 - 8    & 10  & 12  & 16  & 24 & 32 & 48 \\

$N_s$ & 1920 & 1536 & 960 & 448 & 448 & 448 & 64 \\

$T_{m}$ & .9401 & .9793 & 1.0936 & 1.1642 & 1.2059 & 1.3397  
& 1.4084  \\

MCS/$10^6 $ & 3 & 10 & 10 & 10 & 10 & 10  
& 10

\end{tabular}
\end{center}

\label{nsamples}
\end{table}

\begin{figure}[h]
{\epsfig{figure=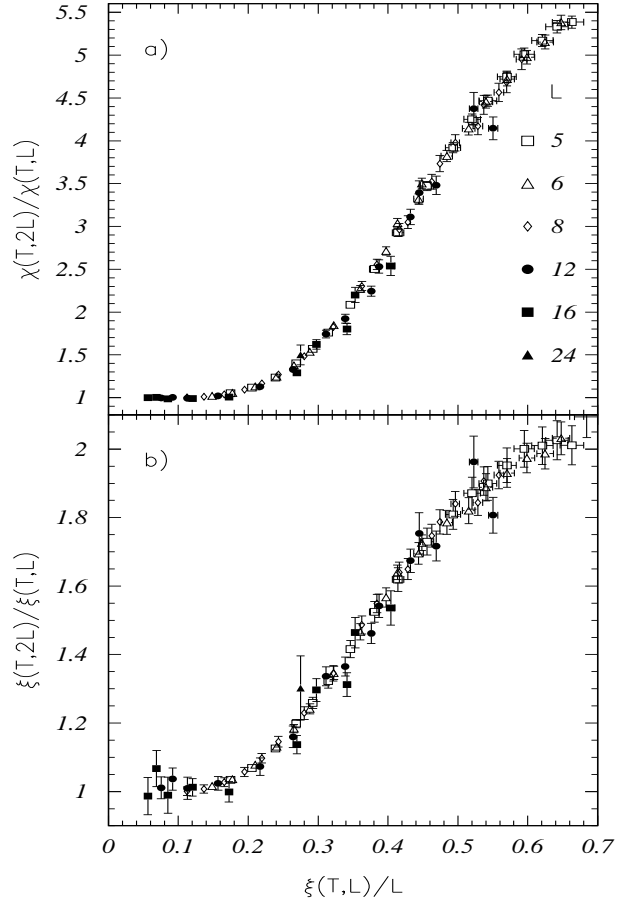,width=10cm,height=12.5cm,clip=}}
\caption{Finite--size--scaling plot with the form in Eq.(\protect\ref{eq2}) 
for a) $\scro=\chi_{SG}$ and b) $\scro=\xi$. Error bars 
(estimated with a jackknife procedure) are one standard deviation.}
\label{fssmethod}
\end{figure}

\paragraph*{FSS analysis ---}
In Fig.~\ref{fssmethod} we show that, within our statistical 
accuracy, the FSS Ansatz \reff{eq2} is well verified for
$\scro=\chi_{SG}$ and $\scro=\xi$. No systematic deviations from the 
curves are detectable, but data at $L=4$, not displayed in Fig.~\ref{fssmethod}, 
are significantly outside the curves for $\xi(T,L)/L \equiv x > 0.2$. 
We verified that other observables, such as the Binder 
ratio, also satisfy Eq.\reff{eq2}. We emphasize that FSS was 
not assumed {\em a priori} 
and that Eq.\reff{eq2} contains no adjustable parameters.
Furthermore, no particular dependence of $\xi$ and 
$\chi_{SG}$ on $T$ was assumed.

We fit the data in Fig.~\ref{fssmethod} to two functions $F_{\chi_{SG}}$, $F_\xi$ 
of the form $F(x)=1+\sum_{i=1,n} a_i \exp(-i/x)$, obtaining
good fits with $n=3$ or $4$  ({\em goodness of fit} 
parameter $Q > 0.9$).
Using $F_{\chi_{SG}}$, $F_\xi$, we then compute 
 $\chi_{SG}(T,\infty)$, $\xi(T,\infty)$  with the iterative 
procedure described above. 
In Table~\ref{tab_extrap} we show that extrapolations from
different $L$ are consistent, providing a test of the method. 
In our final analysis, we take the weighted average of the 
extrapolations from different $L$. 
An implicit assumption of the iterative procedure is 
that the Ansatz \reff{eq2} with a given function $F_\scro$ 
will continue to hold as $L\to\infty$. This assumption could fail if the
system exhibits a crossover at large $L$, as in any FSS analysis.
However, as shown in Table~\ref{tab_extrap}, extrapolations at 
$T=1.4084$ from small $L$  are consistent with data from large $L$, 
which have little or no finite-size effects. We therefore believe
that a crossover is unlikely.

In order to test for systematic errors due to 
corrections to FSS, we repeated the analysis excluding $L=5,6$ from the fits of 
$F_{\chi_{SG}}$, $F_\xi$ and we found that extrapolated data change 
within their error bars. We have a good control 
on the extrapolated data up to $\xi \approx 140$; at lower 
temperatures the statistical errors become
quite large, and the data are more sensitive to the region of high
$x$, where there are few data from large $L$. (The largest $x$ used for the 
extrapolations is $x=0.57$, from $T=1.2059, L=5$). 

\begin{figure}[h]
{\epsfig{figure=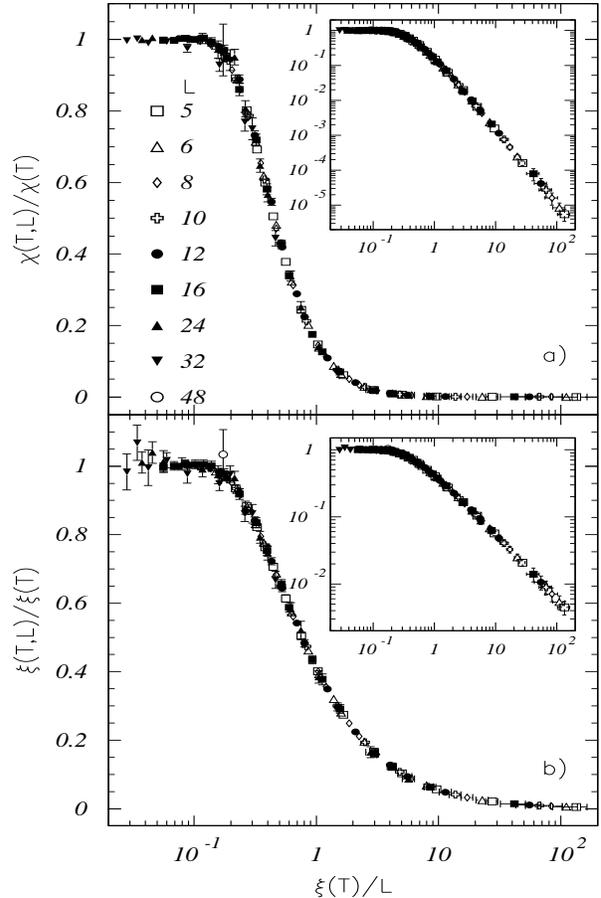,width=10cm,height=12.5cm,clip=}} 
\caption{Finite-size-scaling plot with the form in Eq.\protect{\reff{eq1}}
for a) $\scro=\chi_{SG}$ and b) $\scro=\xi$. The insets represent the same data in a
log-log plot, showing power-law decay for large $\xi/L$.}
\label{fsstheory}
\end{figure}

In Fig.~\ref{fsstheory} we show that with our extrapolated data 
Eq.\reff{eq1} is satisfied remarkably well, providing a further test 
of the method.
If $\scro \sim \xi^{\gamma_\scro/\nu}$ as $\xi \to \infty$, 
then $f_\scro(x)$ in Eq.\reff{eq1} must satisfy 
$f_\scro(x)\sim x^{-\gamma_\scro/\nu}$ as $x\to \infty$. 
As shown in Fig.~\ref{fsstheory} (insets), our curves indeed have a power-law asymptotic decay,
with negative slopes $\gamma/\nu = 2.30 \pm 0.08$ in Fig.~\ref{fsstheory}(a) 
and $\approx 1$ in Fig.~\ref{fsstheory}(b).

We emphasize the  {\em universality} of the scaling functions in 
Fig.~\ref{fssmethod} and \ref{fsstheory}. It would be 
interesting to determine the same functions for different distributions
of the $J_{xy}$, in order to test for possible violations of universality
\cite{campbell1,campbell2}. 

\begin{table}[h]
\begin{center}
\caption{Examples of measured and extrapolated 
values of the correlation length and the spin--glass susceptibility.
See Ref.~\protect\cite{fss_greedy} for how to estimate error bars of extrapolated values.}
\begin{tabular}{r r l l l l} 
\multicolumn{1}{ c}{ $T$}&
\multicolumn{1}{ c}{ $L$}&
\multicolumn{1}{ c}{$\xi(T,L)$ }& 
\multicolumn{1}{ c}{ $ \xi(T,\infty) $ }& 
\multicolumn{1}{ c}{$ \chi_{SG}(T,L)$ }&  
\multicolumn{1}{ c }{$ \chi_{SG}(T,\infty)$}     \\
\hline
1.2059 &  5 &    2.85(7) &  120(60)&     36.1(3) & 1.8(5) $\times 10^5$ \\
       &  6 &    3.42(6) &  150(60)&     55.1(5) & 2.8(8) $\times 10^5$ \\
       &  8 &    4.47(6) &  126(30)&    103(1) & 1.9(5) $\times 10^5$ \\
       & 10 &    5.57(6) &  146(30)&    171(2) & 2.8(8) $\times 10^5$ \\
       & 12 &    6.60(8) &  143(30)&    260(3) & 2.7(8) $\times 10^5$ \\
       & 16 &    8.60(15) & 131(30)&    473(10) & 2.1(8) $\times 10^5$ \\
 &&&&&\\
1.4084 &  5 &    2.28(5) &    8.5(9) &     25.4(2) &    4.3(3) $\times 10^2$\\
       &  6 &    2.66(4) &    8.7(7) &     36.1(4) &    4.6(3) $\times 10^2$\\
       &  8 &    3.33(4) &    8.4(4) &     60.0(6) &    4.3(2) $\times 10^2$\\
       & 10 &    3.94(4) &    8.4(3) &     88.3(1) &    4.3(2) $\times 10^2$\\
       & 12 &    4.51(5) &    8.6(3) &    120(2) &    4.5(2) $\times 10^2$\\
       & 16 &    5.46(8) &    8.6(3) &    178(4) &    4.4(2) $\times 10^2$\\
       & 24 &    6.60(11) &    8.1(2) &    269(6) &    4.2(2) $\times 10^2$\\
       & 32 &    7.16(15) &    7.8(2) &    320(9) &    3.8(2) $\times 10^2$\\
       & 48 &    8.6(6) &    8.9(7) &    404(30) &    4.3(3) $\times 10^2$
\label{tab_extrap}
\end{tabular}
\end{center} 
\end{table}

\paragraph*{Nature of the phase transition ---}
We now compare our extrapolated data with the following scenarios:
(i) a $T_c\neq 0$ continuous phase transition;
(ii) a line of critical points terminating at $T_c \neq 0$, with an
exponential divergence as   $T\to T_c^+$;
(iii) an exponential divergence at $T = 0$. 
The last two scenarios imply a lower critical 
dimension exactly equal to three. 

\begin{figure}[h] 
\centerline{\includegraphics[width=6.0cm,height=8cm,angle=-90]{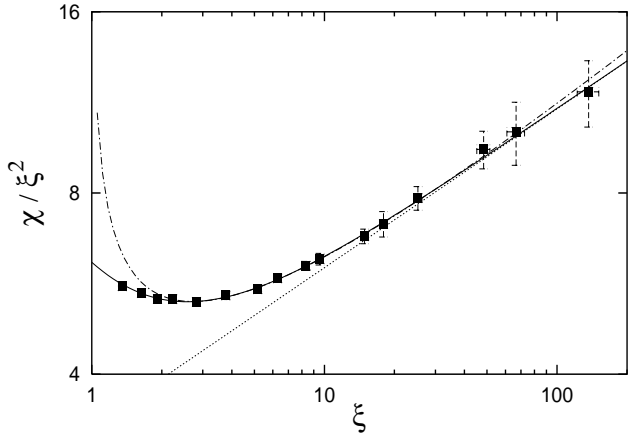}}
\caption{Critical behavior of the infinite volume data. 
The solid line is the best fit to Eq.(\protect{\ref{powerlaw2}}) for $\xi\ge 1.8$, 
the dotted line is the leading term from the same fit, 
the dot-dashed line is the best fit to Eq.(\protect{\ref{logcorr}}) for $\xi\ge 2.2$.}
\label{fitchixi} 
\end{figure}

\begin{figure}[h] 
\centerline{\includegraphics[width=6.0cm,height=8cm,angle=-90]{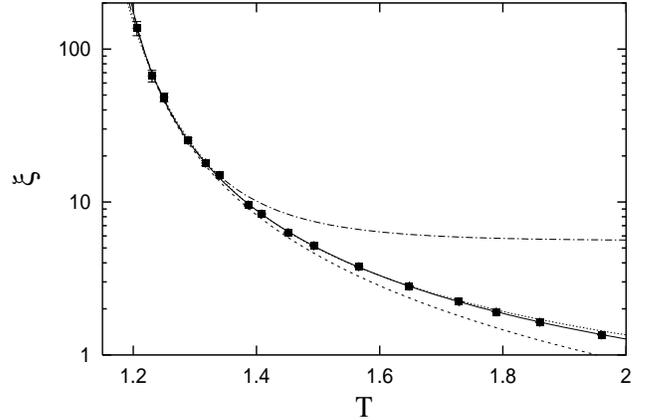}}
\caption{Critical behavior of the infinite volume data. The solid line is 
the best fit to Eq.\protect{\reff{powerlaw1}} for $\xi \ge 1.9$, 
the dotted line is the leading term from the same fit, 
the dashed line is the best fit to Eq.\protect{\reff{KT1}} for $\xi \ge 3.8$,
the dot-dashed line is the best fit to Eq.\protect{\reff{lcd}} for 
$\xi \ge 14$.}
\label{fitxiT} 
\end{figure}

\noindent
(i) We fit our data to 
\beqn
\xi(T) &=& c_\xi \, (T-T_c)^{-\nu} \, \left[1 + a_\xi \, (T-T_c)^\theta \right] 
\label{powerlaw1} \\
\chi_{SG}(\xi) &=& b \, \xi^{2-\eta} \, \left[ 1 + d \, \xi^{-\Delta} \right] 
\label{powerlaw2} 
\eeqn
with fixed correction--to--scaling exponents $\theta$ and $\Delta$
\cite{notalog}. 
In the fit we include data with $\xi \ge \xi_m$, varying
$\xi_m$ in order to test the stability of the fits. 
Without the corrections to scaling
($a_\xi=d=0$), the quality of fits is good for 
$\xi_m > 3 - 4$ ($Q \approx 1$), but fit parameters 
(noticeably $T_c,\nu$ and $\eta$) 
show small {\em systematic} variations with $\xi_m$ in the whole range available. 
Including the corrections, we obtain excellent and stable fits with
$1\le\theta\le 2$ and $1\le\Delta\le 1.5$, the preferred values being 
$\theta=1.4$  ($Q > 0.6$)  and $\Delta=1.3$ ($Q > 0.98$). 
Our estimates  for the fitting parameters 
are $T_c = 1.156 \pm 0.015$, $\nu = 1.8 \pm 0.2$, 
$\eta = -0.26 \pm 0.04$,  $c_\xi = 0.7 \pm 0.2$, 
$a_\xi = 0.5 \pm 0.3$,  $b = 3.3 \pm 0.3$ and $d = 0.9 \pm 0.1$, 
where the errors take into account the uncertainties on $\theta$ and $\Delta$.
We then obtain $\gamma=\nu (2-\eta) = 4.1 \pm 0.5$.
As shown in Fig.~\ref{fitchixi} and \ref{fitxiT}, corrections to scaling are
important for $\xi \le 10$ \cite{fitchi}. Since the fits do not include 
the {\em analytic} corrections to scaling, $\Delta$ and $\theta$ 
should be regarded as ``effective'' exponents.
 For comparison, we quote some estimates from other MC works:
$T_c=1.175\pm 0.025$ \cite{ogielski}, $ 1.11\pm 0.04$ \cite{KY}, 
$ 1.13\pm 0.06$ \cite{janke}, $ 1.19\pm 0.01$ \cite{campbell2};
$\nu=1.3\pm 0.1$ \cite{ogielski}, $ 1.20\pm 0.04$ \cite{on3D}, 
$ 1.7\pm 0.3$ \cite{KY}, $ 2.00 \pm 0.15$ \cite{maparu}, 
$1.33\pm 0.05$ \cite{campbell2};
$\eta=-0.22\pm 0.05$ \cite{ogielski}, $ -0.35\pm 0.05$ \cite{KY}, 
$ -0.30 \pm 0.06$ \cite{maparu}, $  -0.37\pm 0.04$ \cite{janke}, 
$-0.22 \pm 0.02$ \cite{campbell2}
(notice that in Ref.\cite{maparu} a gaussian distribution of the bonds 
was considered).

\noindent
(ii)  We fit our data to 
\beq
\xi(T) = f_\xi \, \exp \bigl( g_\xi / (T-T_c)^\sigma \bigr) \label{KT1}
\eeq
testing the fit stability as above. The fits are
excellent with $\xi_m \ge 1.3$ but, due to strong correlations between
$\sigma$ and $T_c$, the errors on the fit parameters are large. 
For $\xi_m = 1.9$ the best fit gives 
$\sigma = 0.20 \pm 0.05$, $T_c = 1.13 \pm 0.02$, $f_\xi = (1.0 \pm 0.2) \times 10^{-3}$, 
$g_\xi= 7 \pm 2$ ($Q=0.77$). 
Notice, however, that any power-law can be approximated by an exponential 
with sufficiently small $\sigma$. 
For $\xi_m = 3.8$ the best fit (shown in Fig.~\ref{fitxiT}) gives 
$\sigma = 0.5 \pm 0.3$, $T_c = 1.08 \pm 0.04$, $f_\xi = (1.1 \pm 0.8)\times 10^{-1}$, 
$g_\xi = 2.4 \pm 1.5$ ($Q=0.69$).  
The deviations of the data from this fit for $\xi < 3$ are 
consistent with corrections to scaling of $\approx 10\%$.
In general, in the presence of an exponential singularity
we expect multiplicative logarithmic corrections to Eq.\reff{powerlaw2}.
Our data fit well to
\beq
\chi_{SG}(\xi) = b_l \, \xi^{2-\eta_l} \, \left(\log \xi\right)^{r} \label{logcorr}
\eeq
for $\xi_m > 2$, giving
$b_l = 1.30 \pm 0.03$,  $\eta_l = -0.36 \pm 0.03$,   $r = -0.36 \pm 0.06$ ($Q > 0.9$) 
(see also Fig.~\ref{fitchixi}). 

\noindent
(iii) When we fit our data to 
\beq
\xi(T) = f_\xi \, \exp \bigl(g_\xi/T^\sigma \bigr) \label{lcd}
\eeq
we find that $\sigma$ increases continuously with $\xi_m$, from $\sigma \approx 3$ to
$\sigma \approx 9$ \cite{notaxi10}.
Even assuming that $\sigma$ 
stabilizes for higher $\xi$, we believe that a value $\sigma > 9$ is implausibly large.
In fact, Eq.~\reff{lcd} implies a renormalization group (RG) transformation
$dT/dl \propto T^{\sigma+1}$ ($e^l$ being the RG scale factor), while for $T\to 0$
(at the lower critical dimension) we expect $dT/dl = a_2 T^2 + a_3 T^3 + \dots$
($a_2 = 0$ in the phenomenological RG theory of Ref.~\cite{McMillan}).
 
To conclude, we have shown that FSS is verified in the $3D$ Ising spin glass
and that the correlation length diverges at a {\em  finite} temperature. Whether 
this is a conventional continuous phase transition 
(in which case the lower critical dimension is probably close to three) 
or a transition to a line of critical points, is still not known.

We thank A.~Pelissetto, A.P.~Young and O.C.~Martin for useful discussions.
This work was supported by the INFM Parallel Computing
Initiative.


\begin{references}
\bibitem{notadef}
Eq.\reff{corr_len_2mom} is one of the possible definitions giving
$\xi(T,\infty)= \big({1\over 2D} \sum_x |x|^2 G(x)/ \sum_x G(x) \big)^{1\over 2}$,
where $G(x-y)=\la q_x q_y \ra$.
\bibitem{notalog}
In practice the fits are performed using the 
logarithm of both sides of Eqs.~(\protect{\ref{powerlaw1},\ref{powerlaw2}}). 
The same is true for Eqs.~(\protect{\ref{KT1},\ref{logcorr},\ref{lcd}}).
\bibitem{fitchi} Fits to $\chi_{SG}(T) =  c_\chi \, (T-T_c)^{-\gamma} 
\, [1 + a_\chi \, (T-T_c)^\theta ]$ are much less stable, due to stronger
corrections to scaling ($> 100\%$ for $T > 1.5$). For $\xi_m > 9$
we find $T_c=1.14\pm 0.03$, $\gamma=4.4 \pm 0.6, c_\chi = 0.7 \pm 0.5, 
a_\chi = 5 \pm 6$.
\bibitem{notaxi10}
 With $1 \le \xi \le 9.6$ we get $\sigma = 3.1\pm 0.1$, similarly
to Refs.~\cite{ogielski,on3D} where the largest $\xi$ considered was 
$\xi \approx 10$.
\end{references}
\end{document}